# Improving Performance of Relation Extraction Algorithm via Leveled Adversarial PCNN and Database Expansion

Diyah Puspitaningrum
*Department of Computer Science*
*University of Bengkulu*
Bengkulu, Indonesia
diyahpuspitaningrum@gmail.com

*Abstract*—This study introduces database expansion using the Minimum Description Length (MDL) algorithm to expand the database for better relation extraction. Different from other previous relation extraction researches, our method improves system performance by expanding data. The goal of database expansion, together with a robust deep learning classifier, is to diminish wrong labels due to the incomplete or not found nature of relation instances in the relation database (e.g., Freebase). The study uses a deep learning method (Piecewise Convolutional Neural Network or PCNN) as the base classifier of our proposed approach: the leveled adversarial attention neural networks (LATTADV-ATT). In the database expansion process, the semantic entity identification is used to enlarge new instances using the most similar itemsets of the most common patterns of the data to get its pairs of entities. About the deep learning method, the use of attention of selective sentences in PCNN can reduce noisy sentences. Also, the use of adversarial perturbation training is useful to improve the robustness of system performance. The performance even further is improved using a combination of leveled strategy and database expansion. There are two issues: 1) database expansion method: rule generation by allowing step sizes on selected strong semantic of most similar itemsets with aims to find entity pair for generating instances, 2) a better classifier model for relation extraction. Experimental result has shown that the use of the database expansion is beneficial. The MDL database expansion helps improvements in all methods compared to the unexpanded method. The LATTADV-ATT performs as a good classifier with high precision P@100=0.842 (at no expansion). It is even better while implemented on the expansion data with P@100=0.891 (at expansion factor *k*=7).

## I. Introduction

The relation extraction task can be in an unsupervised or supervised domain. In terms of supervised, relation extraction must classify an entity pair to a set of trained relations using documents that contain the entity pair. A binary classification relation extraction system predicts whether or not a given document contains any relation for the entity pair. In a multi-class classification relation extraction system, the system classifies the ontology of a relation of the given document.

The supervised techniques for machine learning require a large amount of training data for learning. Since hand-annotated datasets consume time, distant supervision [6][9][3][12] produces a large amount of training data by aligning knowledge base facts with text. More sophisticated models in recent years have primarily focused on extracting structured information from plain unstructured text using deep learning methods. The use of PCNN in relation extraction has been topics in recent researches [4][5][15][16].

This study proposes a sentence-level of leveled adversarial attention on PCNN for relation extraction tasks on multi-class classification with the help of database expansion using association rules mining algorithm based on the Minimum Description Length principle (KRIMP algorithm). The algorithm itself produces an excellent representative model of a database since the KRIMP mining only the most frequent/interesting patterns that compress the database best. Many researchers have investigated Krimp implementation in many applications, such as in clustering, classification, and recommender system (see [14][13][7][8]). In the Information Retrieval area, for the relation extraction problem, the KRIMP algorithm can be used to grab only the interesting itemsets taken from the database summary called *code table*. Then PCNN LATTADV-ATT (or shortly LATTADV-ATT) uses those itemsets to capture the possible semantic relationship between pair of entities in every itemset. Itemsets are enlarged using *k* topmost similar patterns/itemsets in code table using Jaccard or Cosine similarity until all pairs of entities found.

Our contributions are two folds:

• To demonstrate the use of database expansion through MDL based semantic identification as a novel preprocessing technique in relation extraction task;

• To propose LATTADV, a novel independent deep learning framework for relation extraction classifier. By independent means that the model also can be implemented on any classification tasks.

## II. Methodology

In this section, we first started the problem definition and hypothesis. Then we introduce the preprocessing method (the database expansion method) used in our experiment and the architecture of the LATTADV classifier model.

### A. Problem Definition

Given a dataset of text sentences with entity pair, $X = \{x_1, x_2,\ldots,x_n\}$, and given $R$ is relation labels, our model measures the probability of each relation $r$, where $r \in R$.

### B. Hypothesis

Our hypotheses are: 1) The MDL database expansion is beneficial to improve the classifier system performance. It is because the MDL can capture very well the association between the most common go together pattern itemsets of all the essential and the most frequent words in $X$. The merged of the most similar association itemsets together with entity identification in the expansion process helps in creating new instances that meet the grammatical/semantic rules. 2) The use of leveled architecture is beneficial to improve the performance of a classifier.

*C. Preprocessing: The MDL Database Expansion Method*

As the dataset, we use pre-trained words learned from the New York Times (NYT) corpus 2005-2006 (https://catalog.ldc.upenn.edu/LDC2008T19). The dataset previously used by many researchers in the field of relation extraction [9][3][12][16]. The entities of the dataset were annotated with Stanford NER (https://nlp.stanford.edu/software/CRF-NER.html) and linked to Freebase (www.freebase.com). As the initial/first embedding words were obtained using word2vec (https://code.google.com/archive/p/word2vec/).

As there are many NA values in the dataset, the relation extraction surely is not a trivial task. It is not only about compares the extracted relation instances against Freebase relations automatically but also how to capture and generate the *k* topmost semantic relations between pairs of entities to produce the new expanding sentences. The role of database expansion is to diminish wrong class labels of relation through a semantic understanding of the entity arrangement. Fig. 1 shows the schema of the LATTADV-ATT relation extraction system with database expansion.

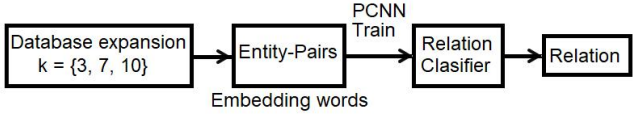

Fig. 1. Overview schema of the leveled adversarial PCNN relation extraction system with database expansion

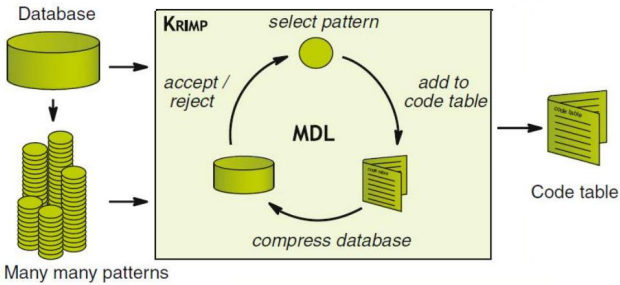

Fig. 2. How KRIMP works

Following is the procedure to expand the database. Given *X* (see Section 2.1), KRIMP, a state-of-the-art algorithm in MDL, will produce a summary of *X* called *code table* (or *CT*). The *CT* is a collection of itemsets. An itemset here denotes the most common words in a dataset of *X* that usually go together. The KRIMP algorithm [14] works as an association rule algorithm, but it only accepts patterns that help compress the dataset *X* optimally. The goal of MDL is to find CT that minimizes the compressed MDL size of the database:

$$L(D, CT) = L(D|CT) + L(CT|D) \quad (1)$$

Assume a *transaction* denotes a set of important words that represent a text sentence $x_i$. The $L(D|CT)$ denotes the size of the encoded database when the CT encodes it, viz. the sum of encoded lengths of all transactions in *X*. The $L(CT|D)$ denotes the size of encoded CT, viz. a sum of the addition of all sizes of code in CT and their length of encoded itemsets. The *size of a code* denotes the length of the code in standard CT (individual item). The *length of the encoded itemset* denotes the length of the encoded itemset throughout the CT. Candidate itemset contains a set of words that frequently go together. It is only inserted into CT if it produces good compression; otherwise, the Krimp will reject the candidate itemset. All the relative frequencies and the compression results also evaluated. MDL only accept CT that minimize $L(D, CT)$ best. See Fig. 2.

We propose a database expansion using itemsets in a code table. Different from Widened Krimp for diverse parallelism [10] that maximizes the sum of minimum distances between pairs of members of the selected subset in CT, in our proposed method, the focus is in the semantic of entity pair. Start from a base itemset; we expand data by combining *k*-topmost itemsets rank by similarity score. Expansion is accepted only if until *k*-topmost itemsets, at least there is one entity pair is presence. A base itemset starts from the first tuple to the last tuple in CT. Every tuple in CT denotes a set of most frequent words that usually go together. Lower minimum support set up in CT leads to more itemsets available in CT.

Consider a set of patterns in CT, $\mathcal{F} = \{F_1, F_2, ..., F_l\}$, the expanded pattern generated from *k*-topmost similar itemsets defined by:

$$F_i^* = \bigcup_{i=1}^{k} \min_{1 \leq j \leq l, i \neq j} \quad (2)$$

Where $F_i, F_j \in \mathcal{F}$ and $\Delta$ denotes text similarity technique, e.g., cosine similarity or Jaccard similarity. The *l* denotes the number of tuples in CT.

There are three kinds of situations about frequent itemsets that KRIMP generates from a sentence:

• Ideal: The *ideal* itemset of $F_i^*$ is a situation where an itemset (a collection of essential keywords of a sentence) has an entity pair inside. If the situation found, the itemset is appended to the database to expand the existing data.

• Half ideal: Another situation is the *half ideal* itemset. This kind of situation is when a pair of entities is found only a half in $F_i^*$, thus creating an incomplete pair. This situation needs an expansion process of *k*-topmost similar pattern generation until $F_i^*$ has at least one entity pair inside. If there is no entity pair found in $F_i^*$ after the combination of *k*-topmost similar itemset, then $F_i^*$ is skipped away, and the base itemset is set up with the next item set.

• Not ideal: This is the situation when no entity found in the itemset. We skip this type of itemset since the expansion method never uses a not ideal code as a base itemset.

In our research, the expansion process focus on three entities: PERSON, LOCATION, and ORGANIZATION. After detecting the availability of the three entities in an itemset in CT (or in a sentence), the possible semantic relations are as follow:

• "/business/company/location": when an entity of ORGANIZATION and entity of LOCATION found in the same sentence. The newly generated instance has a pattern of "X has location Y." Where X is an ORGANIZATION entity, Y is a LOCATION entity.

• "/people/person/place_lived": when an entity of PERSON and entity of LOCATION found in the same sentence. The

newly generated instance has a pattern of "X has location Y." Where X is a PERSON entity, Y is a LOCATION entity.

• "/business/person/company": when an entity of PERSON and entity of ORGANIZATION found in the same sentence. The newly generated instance has a pattern of "X has organization Y." Where X is a PERSON entity, Y is an ORGANIZATION entity.

```
(a). Ideal
  0 1 2 3 : 0 is Entity (PERSON), 2 is Entity (LOCATION)
  0 4 5   : 0 is Entity (PERSON), 4 is Entity (ORGANIZATION)
(b). Half ideal                    (c). Not ideal
  0 5 6 : 0 is Entity (PERSON)       1 3 5 : No Entity is found
  4 7   : 4 is Entity (ORGANIZATION)
  1 2   : 2 is Entity (LOCATION)
CT: 0 1 3 6           Base itemset = 0 1 3 6
    0 3 5 6 (sim=0.75)   k = 3  Extension itemsets:
    1 2 5 6 (sim=0.5)    0136 U 0356 U 035 U 1256 = 012356
    2 5 6 7 (sim=0.25)   0136 U 0356 U 037 U 1256 = 0123567
    0 3 5   (sim=0.57735)
    0 3 7   (sim=0.57735)
    2 5 7   (sim=0)
    4 5 7   (sim=0)
```

Fig. 3. Possible situations when expanding a base itemset

From the above-bulleted items, (X, Y) is a pair of the identified entities, while the sentence in the above items means either in the same itemset or in the same set of combined itemsets ($F^*$). Fig. 3 illustrates the expansion process, while Fig. 4 shows an instance of the generated new sentence.

| Relation | /business/person/company |
|---|---|
| entity1.id | m.0dan05 |
| entity1.name | danay |
| entity2.id | m.0ele14 |
| entity2.name | eleftherotypia |
| $F^*$ | steijn danay eleftherotypia |
| TP | danay has organization eleftherotypia |

Fig. 4. An example of database expansion

*D. The Model*

PCNN models shown to outperform a traditional non-deep learning model in relation extraction [6][3][12]. Based on Convolutional Neural Networks (CNN) [1], Piecewise Convolutional Neural Network (PCNN) [16] has many variants such as attention [4], attention adversarial [15], and maximum adversarial. We selected the PCNN model proposed in [4][15] and built our LATTADV model with two variants based on naive (maximum, or LATTADV-MAX) and sentence-level attention (LATTADV-ATT).

*1) Architecture:* Our proposed method uses three stages of neural networks. Stage one is the PCNN with selective attention of sentences (ATT) followed by two stages of adversarial PCNN training network (ADV) with attention (ATT-ADV). In LATTADV-ATT (Fig. 5), the output of the first stage becomes input in the second stage but with perturbation. The output of the second stage becomes input in the third stage (the final stage) and also with perturbation. The perturbation makes the classifier more robust. In LATTADVMAX, the attention in the second stage substitute by the maximum selector function. Both of the LATTADV methods use the dropout function to prevents units from overfitting and to help efficiently join many different neural network architectures. Dropout is a technique to randomly drop units and their connections from a neural network during training [11]. See [16][15] for PCNN architecture.

*2) Main Components:* Three main components of LATTADV:

*a) Sentence encoder:* Convolutional Neural Network (CNN) is used to build distributed representation of sentence $x_i$, $i = \{1, 2, ..., n\}$.

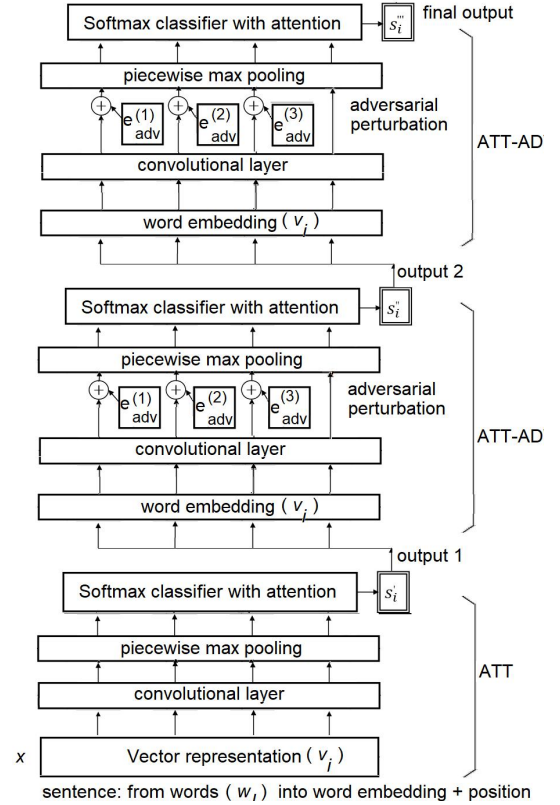

Fig. 5. Architecture of LATTADV-ATT

*b) Selective attention (ATT):* When distributed vector representations of $X$ learned, ATT only gives attention to sentence $s_r$ that sincerely express relation $r$ through computing the weighted average of sentences $s_1, s_2, ..., s_n$, with $s_r$ defined as:

$$s_r = \sum_i \alpha_i^r s_i \quad (3)$$

The softmax function is used to compute the embedding vector of the query vector $q_r$. The $\alpha^r$ denotes the attention weights with relation $r$, $\alpha^r$=softmax(tanh($s_i$)⊤ $q_r$)), see [4]. ATT de-emphasizes noisy sentence. ATT helps to avoid massive false labeling during training ⊤ q(r) and testing. Final loss function for ATT[15]:

$$L_{ATT}(X,\theta) = -\sum_{i=1}^{K} \log P(r \mid x,\theta) \quad (4)$$

The θ denotes parameters, and the $K$ denotes the total number of predefined relation labels $r$. The conditional probability of relation:

$$P(r|x,\theta) = \text{softmax}(A s_r + b) \quad (5)$$

Where $A$ denotes a weighted diagonal representation matrix of relations, $s_r$ denotes attended sentence, and $b$ denotes a bias vector.

*c) Adversarial training (ADV):* Adversarial training first introduced in [2]. In ADV, a small amount of perturbation, $e_{adv}$, is added on word embedding after that adversarial training computes the gradient direction of a loss function with aims to maintain closeness and linearity to input data (viz. by linearizing loss function near the input $X$). Given word embedding of all the words in $X$ called as $V$, $V = \{v_1, v_2, ..., v_n\}$, ADV adds $e_{adv}$ to $V$. The final loss function for ATT-ADV is defined as follows [15]:

$$L_{\text{ATT-ADV}}(X,\theta)=L_{\text{ATT}}(X+e_{adv},\theta) \quad (6)$$

where:

$$e_{adv}=\epsilon g/\|g\| \quad (7)$$

and

$$g = \nabla_v L_{ATT}(X,\hat{\theta}) \quad (8)$$

The $\theta$ denotes all parameters. The $\|g\|$ denotes the norm of gradients over all the words from all the sentences in $X$.

### III. EXPERIMENTS

In this section, we first introduce the dataset we used in our experiment. Then we introduce evaluation metrics and the baselines. After that, we demonstrate the capability of LATTADV as a classifier. Then we discuss the effect of database expansion utilization on the performance of the relation extraction system.

In our LATTADV model, both in ATT and MAX version, for similarity measure, we use cosine and Jaccard similarity, respectively. We tune our experiments with the following parameter settings: sentence embedding size 230, position embedding dimension 5, word embedding size 50, sliding window size 3, maximum of training epochs 60, learning rate 0.5, batch size 160, dropout rate 0.5, weight decay 0.00001, and maximum of relations 53.

TABLE I. STATISTICS OF THE USED NYT DATASET

| Training set | k | #Rel | #Ent-Pair | #Mention | #Sentence |
|---|---|---|---|---|---|
| base | 0 | 53 | 172399 | 293162 | 570088 |
| cos | 3 | 53 | 172499 | 293170 | 570388 |
| jac | 3 | 53 | 172499 | 293170 | 570388 |
| cos | 7 | 53 | 173099 | 293187 | 572990 |
| jac | 7 | 53 | 173099 | 293185 | 572988 |
| cos | 10 | 53 | 173299 | 293185 | 573649 |
| jac | 10 | 53 | 173299 | 293185 | 573649 |

| Testing set | k | #Rel | #Mention | #Sentence |
|---|---|---|---|---|
| base | 0 | 53 | 1950 | 172448 |
| cos | 3 | 53 | 1950 | 172525 |
| jac | 3 | 53 | 1950 | 172525 |
| cos | 7 | 53 | 1950 | 173175 |
| jac | 7 | 53 | 1950 | 173175 |
| cos | 10 | 53 | 1950 | 173341 |
| jac | 10 | 53 | 1950 | 173341 |

Word size=50, position size=5, hidden size=230.
#Rel: number of relation.
#Ent-Pair: number of entity pair.
cos: NYT dataset+expansion with $k$ topmost similar itemsets by Cosine similarity.

TABLE II. STATISTICS OF THE USED NYT DATASET

| Method | k | AUC | Max F1 | P@100 | P@200 | P@300 | Mean |
|---|---|---|---|---|---|---|---|
| LATTADV-ATT | 0 | 0.418 | 0.456 | 0.842 | 0.801 | 0.761 | 0.801 |
| ATT-ADV | 0 | 0.412 | 0.441 | 0.861 | 0.786 | 0.754 | 0.801 |
| LATTADV-MAX | 0 | 0.409 | 0.452 | 0.782 | 0.761 | 0.741 | 0.761 |
| MAX-ADV | 0 | 0.402 | 0.447 | 0.802 | 0.776 | 0.741 | 0.773 |
| ATT | 0 | 0.399 | 0.448 | 0.752 | 0.726 | 0.724 | 0.734 |
| ATT_tanh | 0 | 0.395 | 0.443 | 0.822 | 0.781 | 0.741 | 0.781 |
| RNN-MAX | 0 | 0.395 | 0.441 | 0.762 | 0.756 | 0.734 | 0.751 |

LATTADV-ATT: PCNN + leveled attention and adversarial attention
ATT-ADV: PCNN + attention + adversarial
LATTADV-MAX: PCNN + leveled attention and adversarial + maximum function
MAX-ADV: PCNN + maximum function + adversarial
ATT: PCNN with attention
ATT_tanh: PCNN with attention + tanh activation function
RNN-MAX: recurrent neural network + maximum function

### A. Dataset

The number of the corpus in the NYT dataset (Section 2.3) is 1.8 million news articles and accommodates 53 relation labels. For vocabulary, we use words with frequency ≥100.

In our experiment, the NYT dataset was expanded and split into six datasets as expansion results using Cosine and Jaccard similarity, respectively, with $k=\{3,7,10\}$. The training and testing datasets were generated randomly using cross-validation with a ratio of 80:20. Table I shows the statistics of the NYT extend datasets. Different similarity metrics result in different pattern extensions, even if they have the same $k$ factor.

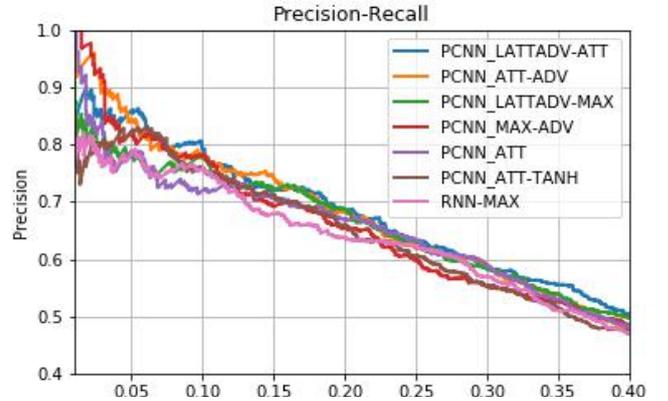

Fig. 6. Architecture of LATTADV-ATT

### B. Evaluation Metrics and Baselines

We measure AUC, max F1, precision-recall, and precision scores at P@100, P@200, P@300 to investigate the performance of all PCNN models with various architectures. Also, the Borda voting score is used to rank the relation extraction method. The higher the score, the better the method. We compare our LATTADV-ATT to several state-of-the-art deep learning methods such as Recurrent based Neural Network (e.g., RNN-MAX) and many PCNN based neural networks, either single-stage or leveled. PCNN ATT [4], PCNN ATT-ADV, and PCNN MAX-ADV [15] is to mention several (see Table II and Table III for other benchmark methods).

## C. Model as a Classifier

Table II and Fig. 6 show the performance of LATTADV-ATT on the original NYT dataset. Figure 6 refers to Table II. It shows that LATTADV-ATT outperforms all other baselines, including the recurrent neural networks. Also, the use of attention selector yields a better result than the maximum selector. Its best performance means that the LATTADV-ATT method can perform as a good classifier.

## D. Effect of Number of Relations in Database Expansion

TABLE III. RESULTS, ORDERED BY BORDA VOTING SCORE

| Alg | k | AUC | Max F1 | P@100 | P@200 | P@300 | Mean | Borda |
|---|---|---|---|---|---|---|---|---|
| M1 | k=0 | 0.399 | 0.448 | 0.753 | 0.726 | 0.724 | 0.734 | 28 |
| M1 | Jac, k=3 | 0.398 | 0.44 | 0.782 | 0.761 | 0.728 | 0.757 | 33 |
| M1 | Jac, k=7 | 0.397 | 0.45 | 0.802 | 0.716 | 0.711 | 0.743 | 35 |
| M1 | Jac, k=10 | 0.404 | 0.438 | 0.822 | 0.796 | 0.734 | 0.784 | 95 |
| M1 | k=0 | 0.399 | 0.448 | 0.753 | 0.726 | 0.724 | 0.734 | 28 |
| M1 | Cos, k=3 | 0.408 | 0.45 | 0.782 | 0.781 | 0.764 | 0.776 | 109 |
| M1 | Cos, k=7 | 0.406 | 0.442 | 0.822 | 0.791 | 0.738 | 0.784 | 100 |
| M1 | Cos, k=10 | 0.4 | 0.44 | 0.822 | 0.751 | 0.73 | 0.767 | 45 |
| M2 | k=0 | 0.412 | 0.442 | 0.861 | 0.786 | 0.754 | 0.801 | 149 |
| M3 | k=0 | 0.402 | 0.447 | 0.802 | 0.776 | 0.741 | 0.773 | 74 |
| M3 | Cos, k=3 | 0.406 | 0.446 | 0.881 | 0.796 | 0.761 | 0.813 | 109 |
| M3 | Cos, k=7 | 0.407 | 0.445 | 0.891 | 0.811 | 0.744 | 0.815 | 100 |
| M3 | Cos, k=10 | 0.399 | 0.438 | 0.861 | 0.781 | 0.728 | 0.79 | 45 |
| M4 | k=0 | 0.41 | 0.452 | 0.782 | 0.761 | 0.741 | 0.761 | 79 |
| M4 | Cos, k=3 | 0.405 | 0.44 | 0.852 | 0.766 | 0.734 | 0.784 | 80 |
| M4 | Cos, k=7 | 0.406 | 0.454 | 0.842 | 0.761 | 0.731 | 0.778 | 88 |
| M4 | Cos, k=10 | 0.405 | 0.44 | 0.852 | 0.766 | 0.734 | 0.784 | 73 |
| M5 | k=0 | 0.418 | 0.456 | 0.842 | 0.801 | 0.761 | 0.801 | 187 |
| M5 | Jac, k=3 | 0.415 | 0.449 | 0.852 | 0.791 | 0.764 | 0.802 | 169 |
| M5 | Jac, k=7 | 0.422 | 0.455 | 0.891 | 0.791 | 0.771 | 0.818 | 207 |
| M5 | Jac, k=10 | 0.416 | 0.451 | 0.861 | 0.796 | 0.751 | 0.803 | 179 |

M1: PCNN ATT; M2: PCNN ATT-ADV; M3: PCNN MAX-ADV;
M4: PCNN LATTADV-MAX; M5: PCNN LATTADV-ATT
Jac: Jaccard similarity; Cos: Cosine similarity

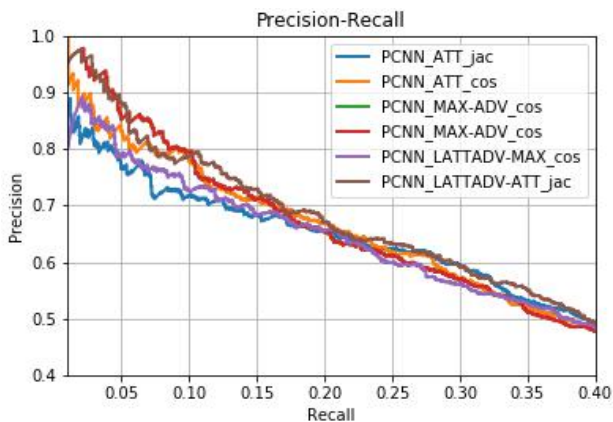

Fig. 7. Precision-Recall of LATTADV-ATT with expansion factor ($k$=7)

Table III and Fig. 7 depicts a list of methods in decrease order by Borda voting methods of all used metrics (AUC, max F1, P@100, P@200, P@300, and mean of precision). Figure 7 refers to Table III. Table III shows that a careful step of expansion factor $k$ can have a significant effect on the improvement of the relation extraction system. Larger $k$ means a more quantity of relations generated in the expanded database (Table I). To obtain better performance, we must test first the appropriate $k$ value carefully because the generated merging itemsets of CT (the $F^*$) may too far away in meaning from the original sentences meaning if the $k$ is too big. As an example, the expansion factor $k$=10 can lead to a long pattern bias of $F^*$. This situation is because the LATTADV-ATT keeps combining the base itemset with all the most similar itemsets found in the CT until at most $k$ itemsets. The best improvement is achieved using LATTADV-ATT with Jaccard similarity expansion at $k$=7 (Table III and Fig. 7).

## E. Effect of The Deep Learning Strategy Setup and The Similarity Metric Used

From Table III, we have proven that the use of the leveled strategy, as well as the attention and the adversarial training in LATTADV-ATT, demonstrates significant improvement in relation extraction system performance, especially against the best state-of-the-art (PCNN ATT-ADV). The LATTADV-ATT outperforms all other methods in all metrics of AUC, maximum F1, and precisions (Table III). From the AUC, the NYT dataset used is imbalanced. Similar to the AUC, the F1 score of the LATTADV-ATT also shows improvement. (The F1 score measurement is to keep a balance between Precision and Recall.) The LATTADV-ATT also improves the precision scores very well. Compared to PCNN ATT-ADV as the best baseline of state-of-the-art methods used, the percentages of the LATTADV-ATT Jaccard $k$=7 improvement are as follows: AUC=2.43%, maximum of F1=2.94%, P@100=3.48%, P@200=0.64%, P@300=2.25%, mean of precisions=2.12%. Also, when compared to the attention network without expansion (PCNN ATT $k$=0), the percentages of the LATTADV-ATT Jaccard $k$=7 improvement are as follows: AUC=5.76%, maximum of F1=1.56%, P@100=18.33%, P@200=8.95%, P@300=6.49%, mean of precisions=11.44%. Therefore the use of the database expansion can improve the relation extraction system's performance. Cosine or Jaccard similarity can be used substitutive.

## IV. CONCLUSIONS

In this paper, we develop a new method for relation extraction task using database expansion and leveled piecewise convolutional neural networks with attention and adversarial training network (LATTADV-ATT). The proposed method can make full use of all informative sentences and serve as an excellent relation classifier on the embedding sentences. Its database expansion feature enlarges the database by generating new sentences that meet the semantic meaning between identified pairs of entities. Experimental results show that: 1) The MDL database expansion is beneficial to improve the performance of the classifier model. This ability is because the minimum description length-based algorithm can sharply capture associations, and the semantic understanding of entity pair helps in generate new expansion sentences. 2) The leveled network with attention and adversarial training strategy in the proposed strategy is beneficial to improve the performance of the PCNN-based relation extraction classifier.